# Cloud Render Farm Services Discovery Using NLP And Ontology Based Knowledge Graph

Ruby Annette (1), Aisha Banu (2), Sharon Priya (2), Subash Chandran (3)

((1) Senior Member IEEE, (2) BSA Crescent Institute of Science and Technology, (3) NEC Corporation of America)

Cloud render farm services are the animation domain specific cloud services Platform-as-a-Service (PaaS) type of cloud services that provides a complete platform to render the animation files. However, identifying the render farm services that is cost effective and also matches the functional requirements that changes for almost every project like the animation software, plug-ins required etc., is a challenge. This research work proposes an ontology-based service discovery engine named RenderSelect for the cloud render farm services. The cloud render farm ontology semantically defines the relationship among the cloud render farm services. The knowledge-based reasoning algorithms namely, the Concept similarity reasoning, Equivalent reasoning and the Numerical similarity reasoning have been applied to determine the similarity among the cloud services. The service discovery engine was evaluated for finding the services under three different scenarios namely a) with help of the ontology, b) without the help of the ontology and c) using a common search engine on the internet. The results show that the proposed service discovery engine which is specifically designed for the cloud render farm services using the ontology performs significantly better than the other two.

**Keywords -** Ontology, Natural Language Processing, NLP, Knowledge Graph, Service Discovery, Similarity Reasoning, Data Mining, Information Retrieval, Cloud Computing, Cloud Rendering, Animation, Cloud Render farm Services

## 1. Introduction

Cloud render farm services are the Platform-as-a-Service (PaaS) type of cloud services that provide their cloud resources and the complete platform to render the animation files [1, 2]. The animation files to be rendered are uploaded onto the Cloud render farm servers using the web interface of the service provider [3, 4]. The uploaded files are assessed by the rendering job queue manager and the render nodes are assigned for completing the rendering job. The updates on the rendering process are displayed in the render management software dashboard and the user has the privilege to monitor, stop or pause the rendering job and pay only for the rendering time for which the cloud render nodes were used. Hence, the cloud render farm services are considered to be a cost-effective alternative for rendering needs in other fields like Fashion designing include Renderingfox, RenderRocket, Rebusfarm etc [5, 6].

Many of our previous work have been focussed on creating a cloud broker service [7,8,9] to aggregate the information about the cloud renderfarms to recommend the right cloud renderfarm services and that let to the realization of the significance of an ontology of cloud renderfarm services to discover and recommend the right cloud renderfarm services. Though many have worked on cloud rendering [10,11,12] and also towards developing an ontology-based service discovery engine for the generic IaaS (Infrastructure-as-a-Service) like Sim KM, et al [13,14,15], no work has considered developing domain specific service discovery engine for cloud render farm services of PaaS (Platform-as-a-Service) type. This research work proposes ontology-based domain specific service discovery engine named RenderSearch for the cloud render farm services of PaaS (Platform-as-a-Service) type. The contributions of this research work include the following: i) This work proposes service discovery engine architecture for domain specific cloud render farm services. (Section ii) Applies the domain specific knowledge gained by developing the cloud render farm services taxonomy to develop a more pruned PaaS (Platform-as-a-Service) specific ontology using the "protege" software that semantically defines the relationship among the cloud render farm services. (Section 3) iii) Identifies and applies three knowledge-based reasoning algorithms namely, the Concept similarity reasoning, Equivalent reasoning and the Numerical similarity reasoning to determine the similarity among the cloud services. (Section 4) iv) Experiments have been carried out using three different search approaches namely: a) without using RenderSearch b) using RenderSearch without Ontology and c) using RenderSearch with ontology and the search results obtained were evaluated using three metrics namely the Precision (P), Recall (R ) and F1. (Section 5) v) Finally, the related works have been discussed; the useful insights gained have been discussed and concluded with scope for further research. (Section 6 and 7)

2. **RenderSearch Service Discovery Engine Architecture:**

Render Search, the proposed cloud render farm services discovery engine architecture is given in Figure 1. Render Search collects the service details from the RaaS (Cloud based Rendering-as-a-Service providers) or the cloud render farm service providers through the Service Profile Collector (SPC) when they register their services using the interface. Whereas the functional requirements details from the user are collected by the requirements collector (RC). Apart from the registered services, the semantic layer is responsible for collecting the details about the non-registered cloud render farm services from the Internet.

The data about the functional requirement that could be met by the cloud render farm services are extracted by scraping their websites. The extracted data were further pruned using the data mining techniques like the data cleaning and stemming techniques etc. The semantic layer also contains the cloud render farm services domain specific ontology developed using the open source "protégé" software [16].

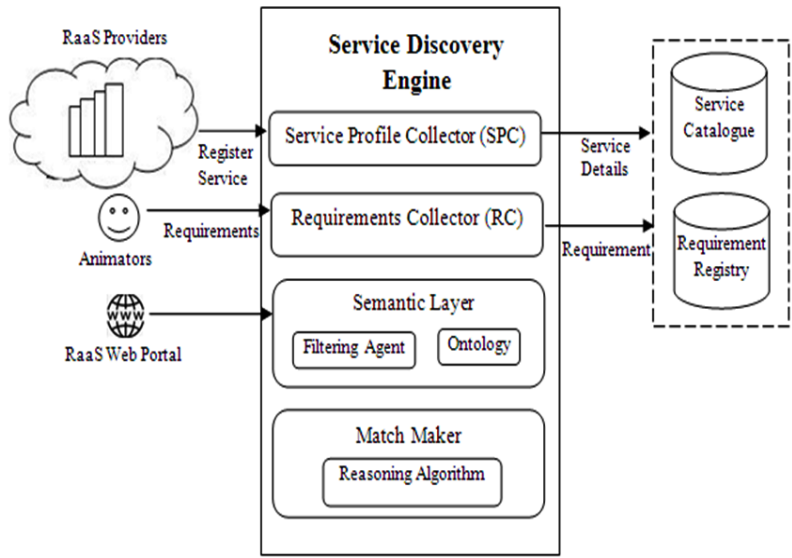

Fig: 1 Proposed service discovery engine Architecture

The Matchmaker module applies three similarity reasoning methods of the knowledge-based filtering techniques on the cloud render farm domain specific ontology to identify the services that could satisfy the functional requirements of the animators. The three similarity reasoning methods used include the Similarity reasoning, Equivalent reasoning and the Numerical reasoning methods. The cloud render farm domain specific ontology and the three reasoning methods are explained in detail in the following section.

3. **Cloud Render Farm Services Domain Knowledge Acquisition:**

**3.1      Developing taxonomy of cloud render farm services:**

Taxonomy is the classification of highly correlated components without any context and ontology on the other hand provides additional context to the classified components [17,18]. A taxonomy may be converted to an ontology to provide more context and knowledge to the knowledge graph developed to make correct information retrieval at times of ambiguity and many works have been done in this direction for cloud services domain [19, 20, 21 ]. The ontology specific to the cloud renderfarm services is not available, however, the domain specific knowledge for which the service discovery engine is built is crucial for the effectiveness of a service discovery engine.  As the foremost step in acquiring the cloud render farm domain specific knowledge and the real time cloud services in the market, were analyzed and the key characteristics of these services were identified and an overall taxonomy of the cloud render farm services was developed.  Later, to enable easy classification and identification of services, the overall taxonomy design proposed was further drilled down to a tree Structured taxonomy. The proposed taxonomy identified seven main characteristics of the cloud render farm services, which includes the Deployment models, Delivery models, User group, License Types, Cost, Operational Transparency and Support as illustrated in detail in the work of Ruby et al [17].

**3.2      Developing Ontology of Cloud Render farm Services**

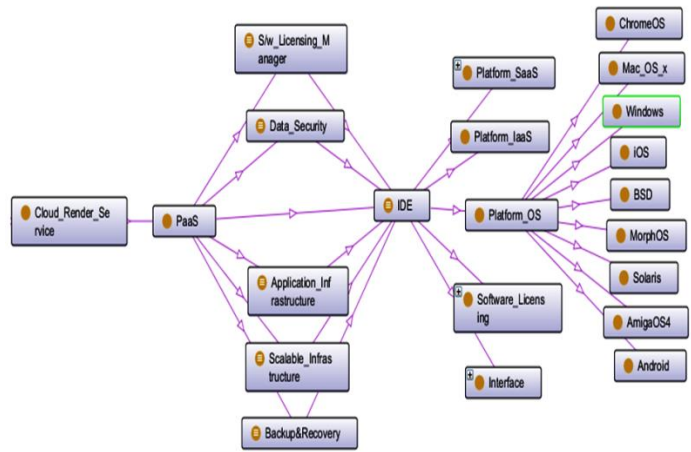

Fig 2: Overall Ontology of PaaS Type Cloud Render farm Services

 Ontologies are more useful compared to taxonomy, as the former defines the most commonly used terminology in the specific domain along with the relationship among them. Whereas, a taxonomy does not define the relationship among the terminologies. Thus, ontologies help to understand the structure of information and the relationship among the terminologies in a domain for both users and software agents. When cloud computing technology bloomed in the market, many focused on developing ontology for the cloud computing domain [22, 23].  Sim KM, et al [13 - 15] developed a cloud service discovery engine based on their proposed ontology for the IaaS type of general cloud services. However, this proposed ontology could not be used to identify the PaaS type of cloud render farm services. Hence, this research work, applies a cloud render farm services specific ontology developed using the protégé tool based on the tree structured taxonomy of cloud render farm services [17].

The overall ontology of the PaaS type of cloud render farm services is given in Figure 2. Whereas, the sub part of the main ontology like the PaaS_ IaaS expands and has the ontology of all the infrastructure related attributes like the CPU unit type, memory, Virtual Machine etc. The PaaS_SaaS also expands and gives the detailed ontology of the software as a service layer which includes the required rendering job management software and a detailed ontology of the software required for rendering an animation file like the 3D animation software, render engine software, plug-in required etc though not shown explicitly in the figure..

## 4. KNOWLEDGE BASED FILTERING USING SIMILARITY REASONING:

The similarity reasoning methods of the knowledge based filtering techniques have been applied in this work. Three different reasoning algorithms based on the type of data to be matched have been identified as the Concept similarity reasoning, Equivalent reasoning and the Numerical similarity reasoning [20]. The concept similarity reasoning method has been applied to measure the degree of commonality between two concepts x and y. Whereas, to determine the similarity between two sibling concepts the equivalent reasoning method has been applied. The equivalent reasoning method uses the label values to calculate the similarity between two sibling concepts. The three similarity reasoning methods are discussed in detail below.

### 4.1 Concept Based Similarity Reasoning

The degree of commonality between the two concepts, namely the x and y are measured using the Formula 1 in the Concept based similarity reasoning method.

$$Sim(x,y) = \rho \frac{|\alpha(x) \cap \alpha(y)|}{|\alpha(x)|} + (1-\rho) \frac{|\alpha(x) \cap \alpha(y)|}{|\alpha(y)|} \quad \ldots (1)$$

Where,

α (x) - set of nodes that can be reached upwards from x including x.

α (y) - set of nodes that can be reached upwards from y including y.

ρ - determines the generalization's degree of influence.

α (x) ∩ α (y) - number of reachable nodes that is shared by x and y.

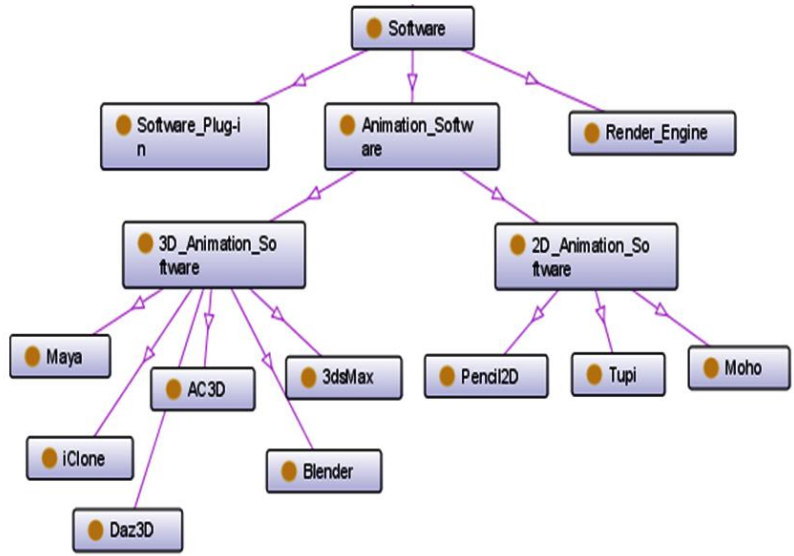

**Figure 3 Ontology of Animation Software**

**Example: 1**

Calculation of Similarity between the 3DsMax and AC3D nodes

ρ = 0.5 (Assumed)

α (3DsMax) = x = 4

α(AC3D) = y = 4

α(x) ∩ α(y) = 3

Sim(x,y) = [ 0.5 (3/4) + 0.5 (3/4)] = 0.75

As an example, let us consider the ontology of 'Animation Software' given in Figure 3 and calculate Sim(x,y), the similarity between two different nodes. Consider the two nodes of the 3D animation software namely, 3DsMax and AC3D, based on the ontology of the software, the set of nodes that can be reached upwards from α (3DsMax), including α (3DsMax) is 4 and that of α(AC3D) is 4. The number of reachable nodes that are shared by both 3DsMax and AC3D is represented as α (x) ∩ α (y) and its value is 3. Hence, the value of Sim(x,y) that represents the similarity between the two is calculated to be 0.75 as shown in example 1.

As the second example, let us calculate the similarity between the '3DsMax' and 'Pencil2D' that belong to two different software categories like the 3D animation and 3D modeling software respectively. In the example 2, α (3DsMax) represents the set of nodes that could be reached upwards from α (3DsMax), including α (3DsMax) and its value is calculated to be 4. Similarly, the Sim(x,y) value of α(Pencil2D) that represents the set of nodes that could be reached upwards from α(Pencil2D), including α(Pencil2D) is also calculated to be 4. However, the number of reachable nodes shared by both '3DsMax' and 'Pencil2D' is represented as α (x) ∩ α (y) and its value is calculated to be 2. Hence, using the formulae given above in Formula 1, the similarity between the two nodes indicated as Sim(x,y) is calculated to be 0.5.

**Example: 2**

Similarity between 3DsMax and Pencil2D

α(3DsMax) = x = 4

α(Pencil2D) = y = 4

α(x) ∩ α(y) = 2

Sim(x,y) = [ 0.5 (2/4) + 0.5 (2/4) ] = 0.5

Only if the similarity value is greater than 0.5, the two concepts compared are considered to be part of similar concepts, otherwise they are not considered to be part of similar concepts. From the concept-based similarity values obtained from example 1 and 2, it is clear that the similarity between the two nodes in example 1 is 7.5, and is above the set threshold value of 0.5, hence 3DsMax' and 'AC3D' are considered to belong to the same concept. Whereas in example 2, since the similarity between '3DsMax' and 'Pencil2D' is calculated to be only 0.5, since it is not higher than the threshold value, they are not considered to be long to the same concept. If the Similarity value of both are the same, then they are assigned as sibling nodes and the equivalent similarity reasoning is performed to match them further. For example if both the user requirement and the searched node are both Maya, then their Similarity score will be 1. In this case the nodes containing the value as Maya are assigned as sibling nodes and the equivalent similarity reasoning is performed further to match their software versions as given in detail below.

### 4.2 Equivalent reasoning

The sibling nodes are identified using the concept-based reasoning methods as discussed in the above step. The similarity between the sibling nodes is determined based on their label values using the Formula 2 of the Equivalent similarity reasoning method as given below.

$$\widehat{Sim}(x,y) = Sim(x,y) + \frac{\left(0.8^{|c_1-c_2|}\right)}{10} \quad \ldots (2)$$

Where,

Sim(x,y) – similarity value.

C1 – represents the label value of concept x

C2 – represents the label value of concept y

The ontology of the software versions of the Maya is given in Figure 4. The similarity in the software versions of the Maya software can be determined based on the similarity of their label values. For example, in Figure 4, the labeling of the software version starts from the oldest version as 1 (Maya 1.0 version), whereas the latest version (Maya 4.0.2) has been labeled as 12. Further, let C1 represent the label value of concept x (Maya 3.0) is 7 and C2 represent the label value of concept y (Maya 4.0.2). Using the formula 4.2, the label value C1 of concept x (Maya 3.0) is calculated to be 7 and that of C2 of concept y (Maya 4.0.2) as 12. Assuming the Sim(x,y) value as 0.5, the equivalent similarity value between Maya 3.0 and Maya 4.0.2 is calculated as 0.5327 as given below.

Let Sim(x,y) = 0.5, then

'Sim (Maya 3.0, Maya 4.0.2) = 0.5+ 0.8 $^{|7-12|}$ / 10 = 0.5327

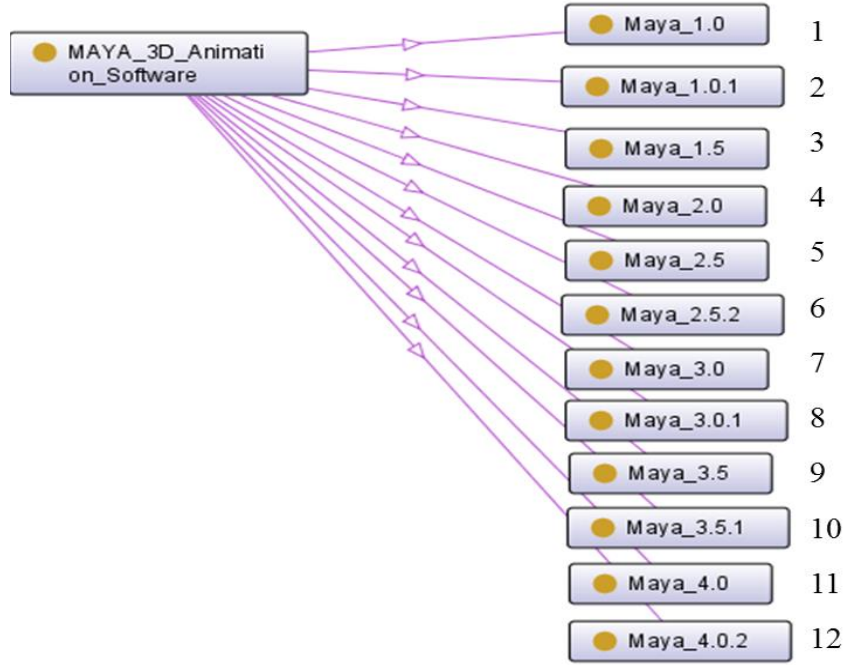

Figure 4: Ontology of Maya 3D Animation Software

Similarly, if C1 represents the label value of concept x (Maya 1.0) and C2 represents the label value of concept y (Maya 4.0.2). Then using the formulae given above, the value of C1 is calculated to be 1 and that of C2 to be 12. The similarity between Maya 1.0 and Maya 4.0.2 is also calculated to be 0.5085 as given below.

'Sim (Maya 1.0 & Maya 4.0.2) = 0.5 + 0.8 $^{|1-12|}$ / 10 = 0.5085

From the given example, it is clear that the Maya 4.0.2 is more similar to Maya 3.0 than Maya 1.0.

### 4.3 Numerical reasoning

Numerical reasoning has been applied to calculate the similarity between two concepts of numeric data type in the same domain using the formula given below:

$$Sim(a,b,c) = 1 - \left| \frac{a-b}{Max_c - Min_c} \right| \quad \ldots (3)$$

Where, 'a' indicates the first numeric value & 'b' indicates the second numeric value and 'c' indicates the concept name. In this work, the numerical reasoning method has been applied to calculate the similarity of the numeric value-based attributes. For example, suppose we would like to filter the cloud render farm services that charges 3.5$ as the render node cost, then a numerical similarity reasoning method can be applied as the render node cost is a numeric value-based attribute. Let us assume that 'a'

indicates the desired render node cost, 'b' indicates the numeric value that has to be compared with the desired value 'a', whereas, 'c' indicates the concept name i.e., the render node cost. Let 'Max$_c$' and 'Min$_c$' indicate the maximum and minimum values of the render node cost.

Case: 1

As case 1, let us calculate the similarity between the desired render node cost of 3.5$ and the actual render node cost of 2.5$. If the assumed value of 'Max$_c$' is 6.0, 'Min$_c$' is 1.0 then, the calculated similarity value is 0.80 as shown below.

Sim (3.5, 2.5, Rendernode Cost),

$$= 1 - \left|\frac{3.5 - 2.5}{6.0 - 1.0}\right| = 0.80$$

Sim (3.5, 5.5, Rendernode Cost),

$$= 1 - \left|\frac{3.5 - 5.5}{9.0 - 2.0}\right| = 0.71$$

Thus, the numerical similarity reasoning method can be used to filter the concepts of numeric type. Finally, the similarity score obtained using all the three methods is aggregated using the Aggregate Sim (R) method and the services are ranked based on the aggregated similarity score in decreasing order. The cloud render farm service with the highest similarity value is recommended first followed by the others in the descending order of their aggregated similarity values.

| Functional Requirement Attribute | Values |
|---|---|
| Compute Unit Type | CPU |
| Supported S/W License Fee | Provided and Fee included |
| Job Management S/W Requirement | Plugin |
| S/W Support Requirements | Cinema 4D |
| Render Engine S/W Requirements | Mental Ray |
| Plugin Requirements | FumeFX |
| OS Requirements | Mac |
| Render Node Cost | 0.60 $ per core per hour |

Table 1: Example of Functional requirements

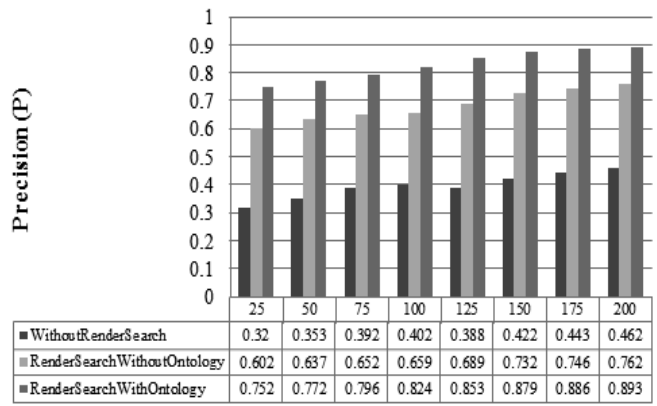

Precision (P) Values for Various Search Approaches

| | 25 | 50 | 75 | 100 | 125 | 150 | 175 | 200 |
|---|---|---|---|---|---|---|---|---|
| WithoutRenderSearch | 0.32 | 0.353 | 0.392 | 0.402 | 0.388 | 0.422 | 0.443 | 0.462 |
| RenderSearchWithoutOntology | 0.602 | 0.637 | 0.652 | 0.659 | 0.689 | 0.732 | 0.746 | 0.762 |
| RenderSearchWithOntology | 0.752 | 0.772 | 0.796 | 0.824 | 0.853 | 0.879 | 0.886 | 0.893 |

Figure: 5 Precision (P) Values

## 5. EXPERIMENTS AND EVALUATION OF THE SERVICE DISCOVERY ENGINE

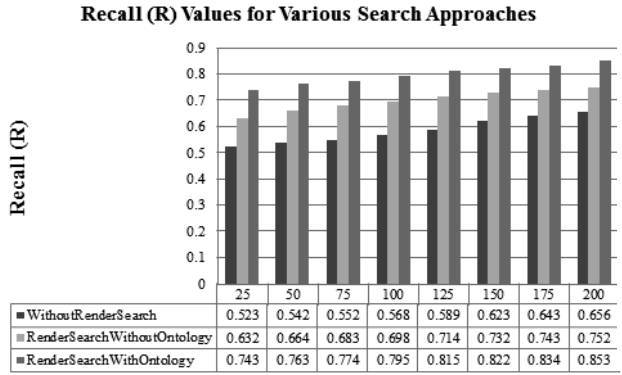 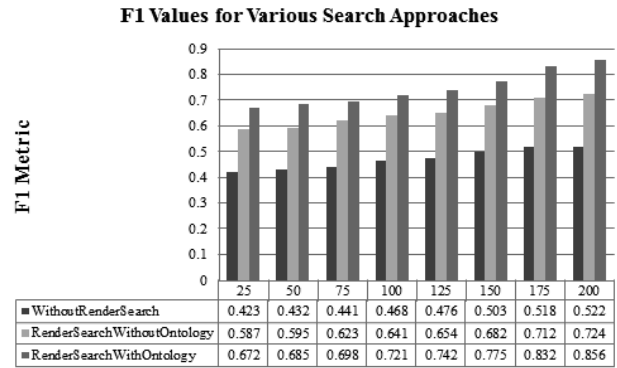

Figure: 6 Recall (R) Values        Figure: 7 F1 Values

A number of experiments were carried out to evaluate the proposed RenderSearch Interface in discovering the cloud render farm services. Three approaches were applied namely: a) without using RenderSearch b) using RenderSearch without Ontology and c) using RenderSearch with ontology to search and discover the cloud render farm services that satisfy the selected functional requirements of the users in a stipulated timeline. The five functional requirements selected include, a) Compute unit, b) Software Supported, c) render Engine Software, d) Plugin, e) Operating System. The test sets were created by varying the values of these five requirements and experiments were carried out to discover services using the three approaches. Consider that an animator prefers to use the Cloud Rendering-as-a-Service (RaaS) to render an animation scene. His need is to identify some cloud render farm services, which would satisfy the functional requirements required to render the animation file and compare their cost. An example of the functional requirements to render the animation file is given in Table 1.

The user enters the functional requirements in the RenderSearch to identify the services. The query is matched by applying the reasoning algorithms onto the cloud render farm services ontology. A list of services that match the functional requirements of the user, arranged in decreasing order based on the aggregated similarity score is provided to the user by the proposed RenderSelect service discovery engine. The search results were evaluated based on three selected metrics namely the Precision (P), Recall (R), F1 metric and the graphs are given in Figure (5, 6 and 7) respectively.

The Precision metric (P) calculates the ratio of the True positives (relevant retrieved results) to the sum of True positives and False positives (overall retrieved results). The second metric considered for evaluation is the Recall (R) metric, it is defined as the ratio of the relevant retrieved items (True positives) to the total of relevant items in the database (True positives + False negatives). The third metric is the F1 Metric, which is based on two other calculated metrics, the Precision (P) and the Recall (R) values. The F1 score is considered best when it is 1 and as worst when the score is 0.

The results obtained for filtering services using the service discovery engine in three different situations namely: a) without using RenderSearch b) using RenderSearch without Ontology and c) using RenderSearch with ontology engine without the ontology were evaluated using three metrics namely the Precision (P), Recall (R ) and F1 and the graphs are given in the Figures (5, 6 and 7) respectively. It is observed from Figure 5, that the Precision (P) of the search results is high at 0.893, when the RenderSearch with ontology approach is used compared to not using the ontology (0.762). However, the Precision value (P) is the least, when the RenderSearch is not used (0.462). Comparing the Recall (R) metric values of the search results in Figure 6, it is also high at 0.853, when the RenderSearch with ontology approach is used compared to not using the ontology at 0.739, the Recall (R) metric values is the least (0.656) when the RenderSearch is not used for discovering the services. Comparing the F1 metric values of the three approaches applied in Figure 7, it can be observed that the F1 value is the least at 0.522 when the RenderSearch is not used for discovering the services and is the highest at 0.856, when the RenderSearch with ontology approach is used. However, when the search is made using the RenderSearch but without using the ontology, the F1 value is at only 0.724.

6. RELATED WORK

Many research works have focused on searching and selecting the services based on the search criteria. The work of Tran et al [24] on the QoS ontology and its QoS-based ranking algorithm for Web services was an important contribution that led to further research on ranking the cloud services for selection by many researchers like Garg et al [25] on SMI attributes based cloud services ranking, Wang et al [26] work on cloud model for service selection etc. Inspired by these work, we have worked on the problem of classifying, ranking and recommending cloud renderfarm services [27-30]. During these works we felt the need for an ontology specific to the cloud renderfarm services and that inspired this work on ontology and similarity reasoning for identifying the right cloud services.

Hussain et al [31] has worked towards multi-criteria cloud service selection, Sundareswaran et al [32] have worked on a brokerage-based approach for cloud service selection. Many research works have also focused on developing ontology for the cloud services. For example, The works of [33, 34] have focused towards developing taxonomy and a unified ontology of cloud computing services; Fortis et al [35] have worked towards developing ontology for cloud services. Some research works have focused towards developing ontology based service discovery engines for the generic IaaS (Infrastructure-as-a-Service) like Sim KM, et al [13-15] .

However, no work has considered developing domain specific service discovery engine for cloud render farm services of PaaS (Platform-as-a-Service) type based on our literature survey. This research work focuses on bridging this gap and proposes an ontology based domain specific service discovery engine for the cloud render farm services of PaaS (Platform-as-a-Service) type as explained in the previous sections.

7. CONCLUSION AND FUTURE WORK

This proposed work has applied three different similarity reasoning methods of the knowledge-based filtering approach to search and discover the cloud render farm services that could satisfy the functional requirements of the users. The proposed service discovery engine uses the ontology to calculate the similarity between the nodes and identifies the services that can satisfy the functional requirements of the users. The search results were evaluated using three different approaches namely: a) without using RenderSearch b) using RenderSearch without Ontology and c) using RenderSearch with ontology. From the evaluation of the search results obtained, we could find that the RenderSearch without Ontology approach performs better than the other two approaches. In the future, the service discovery engine will be further extended as a recommendation engine that would study the user preferences and recommend services based on the user preferences.